# AI based design of 2D-material integrated optical polarizers

Rong Wang [1, 2, †], Di Jin [1, 3, †], Junkai Hu [1, 2, 3], Wenbo Liu [1, 4, 5], Yuning Zhang [6], Irfan H. Abidi [3, 7], Sumeet Walia [3, 7], Baohua Jia [3, 4, 5], Duan Huang [8, *], Jiayang Wu [1, 3, *], and David J. Moss [1, 3, *]

[1] Optical Sciences Centre, Swinburne University of Technology, Hawthorn Victoria 3122, Australia

[2] School of Automation, Central South University, Changsha 410083, China

[3] Australian Research Council (ARC) Centre of Excellence in Optical Microcombs for Breakthrough Science (COMBS)

[4] School of Science, RMIT University, Melbourne, Victoria 3000, Australia

[5] Australian Research Council (ARC) Industrial Transformation Training Centre in Surface Engineering for Advanced Materials (SEAM), RMIT University, Melbourne, Victoria 3000, Australia

[6] School of Physics, Peking University, Beijing, 100871, China

[7] Centre for Opto-electronic Materials and Sensors (COMAS), School of Engineering, RMIT University, 124 La Trobe Street, Melbourne 3000, Australia

[8] School of Electronic Information, Central South University, Changsha 410083, China

[†]These authors have equal contributions to this work.

E-mails: duanhuang@csu.edu.cn (Duan Huang), jiayangwu@swin.edu.au (Jiayang Wu), dmoss@swin.edu.au (David J. Moss)

**ABSTRACT**

On-chip integration of highly anisotropic two-dimensional (2D) materials offers new opportunities for realizing high-performance polarization-selective devices. Obtaining optimized designs for such devices requires extensively sweeping large parameter spaces, which in conventional approaches relies on massive mode simulations that demand considerable computational resources. Here, we address this limitation by developing a machine learning (ML) model based on fully connected neural networks (FCNNs). Trained by using mode simulation results for low-resolution structural parameters, the FCNN model can accurately predict polarizer figures of merits (FOMs) for high-resolution parameters and rapidly map the global variation trend across the entire parameter space. We test the performance of the FCNN model using two types of polarizers with 2D graphene oxide (GO) and molybdenum disulfide ($MoS_2$). Results show that, compared to conventional mode simulation approach, our approach can not only reduce the overall computing time by about 4 orders of magnitude, but also achieve highly accurate FOM predictions with an average deviation of less than 0.04. In addition, the measured FOM values for the fabricated devices show good agreement with the predicted ones, with discrepancies remaining below 0.2. These results validate artificial intelligence (AI) as an effective approach for designing and optimizing 2D-material-based optical polarizers with high efficiency.

**INTRODUCTION**

The rapid advancement in artificial intelligence (AI) technology is revolutionizing the modeling and engineering of optical devices [1-4]. Unlike traditional simulation methods that rely on iterative Maxwell equations solving, AI can offer substantial improvement in computing efficiency by directly mapping the relationships between structural parameters and optical responses [5-8]. This transformative approach is broadly applicable, with particular strength in addressing complex design challenges and optimizing sophisticated devices [1, 2, 9-11]. To date, the use of AI in optical device design has yielded substantial success in developing a variety of functional devices such as metasurfaces [8, 12-15], nonlinear optical devices [16, 17], electro-optic modulators [18], photodetectors [19-21], and quantum optical devices [22-24].

Optical polarizers are essential building blocks for selecting and controlling light polarization states in optical systems [25-28]. Recently, two-dimensional (2D) materials with strong anisotropic light absorption and broadband response have been utilized to implement optical polarizers, offering high device performance and novel features beyond what traditional bulk materials can achieve [25, 26, 29-31]. The performance of 2D-material-based optical polarizers is highly dependent on their device structural parameters [25, 30, 32, 33]. Optimizing the design for these devices typically requires extensive sweeps across large parameter spaces. Conventional methods, which rely on mode simulations via software, are computationally demanding, particularly given that accurate simulations of mode profiles for devices with 2D materials require extremely fine meshing [34-36].

In this work, we report the first machine learning (ML) model based on fully connected neural networks (FCNNs) for the design and optimization of 2D-material-based optical polarizers, significantly improving computing efficiency and maintaining high prediction accuracy. Trained on mode simulation results for low-resolution structural parameters, the FCNN model can accurately predict polarizer figures of merit (FOM) for high-resolution parameters and rapidly map the global FOM variation trend by sweeping the full parameter space. Two types of optical polarizers with 2D graphene oxide (GO) and molybdenum disulfide ($MoS_2$) films are employed to test the performance of the FCNN model. Results show that our ML method can finish sweeping the full parameter space in 25 – 35 s, in contrast to several months as required for mode simulations, and provides accurate FOM predictions with an average deviation (*AD*) of less than 0.04 relative to mode simulations. In addition, the measured FOM values for the fabricated polarizers agree well with those predicted by the FCNN model, with discrepancies below 0.2. These results confirm the effectiveness of AI as an efficient tool for the design and optimization of 2D-material-based optical polarizers.

## RESULTS AND DISCUSSION

Fig. 1a shows the schematic of a 2D-material-based integrated waveguide polarizer, where a silicon (Si) photonic waveguide is coated with an atomically thin 2D material film, such as GO or $MoS_2$. The right panel of Fig. 1a depicts a schematic cross section of the hybrid waveguide, where *W* and *H* represent the width and height of the Si waveguide, respectively, and *n*, *k*, and *d* represent the refractive index, extinction coefficient, and thickness of the 2D material film, respectively. These structural parameters are critical for designing the polarizer and optimizing its performance.

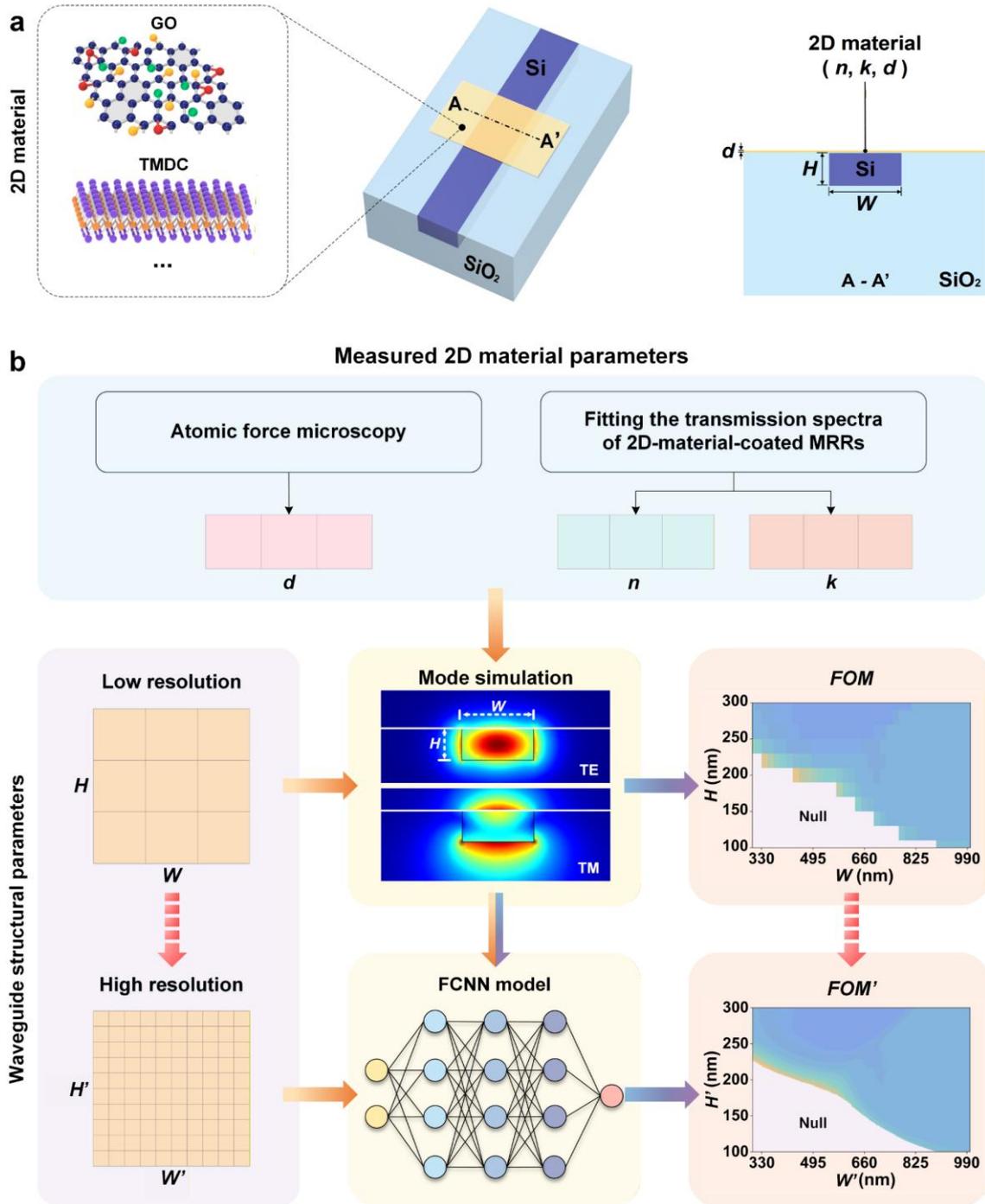

**Fig. 1 | a,** Schematic illustration of a 2D-material-based integrated waveguide polarizer consisting of a silicon (Si) photonic waveguide coated with a 2D material film such as graphene oxide (GO) and molybdenum disulfide (MoS$_2$). The right panel shows the cross-sectional view of the hybrid waveguide, where *W* and *H* denote the width and height of the Si waveguide, respectively, and *n*, *k*, and *d* represent the refractive index, extinction coefficient, and thickness of the 2D material film, respectively. **b,** Schematic illustration of a fully connected neural network (FCNN) model used to predict polarizer figures of merit (*FOM'*) for high-resolution structural parameters (*W'*, *H'*) based on mode simulations with low-resolution structural parameters (*W*, *H*).

Recent studies have revealed that 2D materials such as graphene [26, 29, 31, 37, 38], GO [30, 32, 33, 39, 40], transition metal dichalcogenides (TMDCs) [41-45], and MXenes [46, 47] exhibit strong anisotropy in their light absorption, with light propagating in the in-plane direction showing significantly stronger absorption than that in the out-of-plane direction. Due to the interaction between the evanescent field from Si waveguide and the 2D material film with strong anisotropy in its light absorption, the hybrid waveguide in Fig. 1a exhibits significantly stronger light absorption for transverse electric (TE, in-plane) polarization compared to transverse magnetic (TM, out-of-plane) polarization. This enables it to function as an effective TM-pass optical polarizer. In addition, 2D materials can exhibit strong anisotropy in their light absorption over a broad spectral range spanning from visible to infrared wavelengths [25]. This wide bandwidth provides a significant advantage for 2D-material-based integrated waveguide polarizers over conventional bulk Si photonic polarizers, which typically have limited operating bandwidths of less than 100 nm [26, 29]. In the following discussion, we use 2D GO and $MoS_2$ (a typical TMDC) films on a Si photonic platform as examples to demonstrate the effectiveness of our FCNN-based model. This choice is motivated by Si's dominance in integrated photonic devices and our extensive prior research on these 2D materials [30, 42, 48-50]. However, the operation principle is not limited to these 2D materials or Si platform and can be readily extended to other 2D materials (*e.g.*, graphene and MXene) and integrated platforms (*e.g.*, silicon nitride [51] and lithium niobate [52]).

Fig. 1b illustrates the process flow for using a machine learning approach to predict polarizer figures of merit (*FOM'*) for high-resolution structural parameters (*W'*, *H'*) based on mode simulations with low-resolution parameters (*W*, *H*). It mainly includes three steps.

First, the 2D material parameters (*n*, *k*, *d*) are experimentally measured, which will be used for mode simulations in the next step. For instance, the 2D material film thickness can be characterized by using atomic force microscopy (AFM) [48]. The TE- and TM-polarized refractive indices ($n_{TE}$, $n_{TM}$) and extinction coefficients ($k_{TE}$, $k_{TM}$) of the 2D material film can be obtained by fitting the transmission spectra of 2D-material-coated microring resonators (MRRs) [53, 54] based on the scattering matrix method [55, 56], which has been widely adopted, with representative works in Refs. [30, 32]. Generally, by comparing the resonance wavelength shifts of the same MRR before and after coating 2D materials and calculating the relative phase shift, the refractive index of the coated 2D material can be extracted. On the other hand, the round-trip loss of a MRR can also be obtained by fitting its transmission spectrum, and the extinction coefficients of 2D materials can be extracted by comparing the round-trip losses of the same MRR before and after coating 2D materials. The main uncertainties in the extracted $n$, $k$ arise from spectral fitting errors and non-uniformity of the 2D material films, which can be mitigated by averaging results over multiple resonances and devices with different coating lengths.

Second, by using the measured 2D material parameters, mode simulations are performed for the hybrid waveguides with low-resolution ($W$, $H$) sets. For small values of $W$ or $H$, the corresponding TE or TM modes may fail to converge, indicating that these dimensions meet the mode cut-off condition and the modes cannot physically exist. In such cases, the corresponding ($W$, $H$) is recorded as 'Null'. When both TE and TM modes are converged, the power propagation losses (dB/cm) of the hybrid waveguide can be calculated by [53, 54]

$$PL_{TE} = -10 \cdot \log_{10} \left\{ \left[ exp(-2\pi \cdot k_{TE,\,eff} \cdot L / \lambda) \right]^2 \right\} \quad (1)$$

$$PL_{TM} = -10 \cdot \log_{10} \left\{ \left[ exp(-2\pi \cdot k_{TM,\,eff} \cdot L / \lambda) \right]^2 \right\} \quad (2)$$

where $k_{TE,\,eff}$ and $k_{TM,\,eff}$ are the imaginary parts of the effective refractive indices for the TE and TM modes, respectively, $L$ = 1 cm is the waveguide length, and $\lambda$ is the light wavelength. Based on Eqs. (1) and (2), the polarizer figure of merit (*FOM*) can be further calculated by [30]

$$FOM = PDL / EIL = (PL_{TE} - PL_{TM}) / PL_{TM} \qquad (3)$$

where *PDL* is the power dependent loss defined as the difference between the TE- and TM-polarized insertion losses, and it has been widely employed to quantify the polarization selectivity of optical polarizers [25, 32, 33]. *EIL* is the minimum excess insertion loss induced by the 2D film, which equals to the excess insertion loss for TM polarization. In 2D-material-based optical polarizers, the 2D material films not only provide high polarization selectivity but also introduce excess insertion loss. Therefore, the *FOM* defined in Eq. (3), which quantifies the trade-off between these two factors, is commonly used to assess the performance of 2D-material-based optical polarizers [25, 29, 30].

Finally, the recorded 'Null' for (*W*, *H*) corresponding to non-converged modes and the calculated *FOM* values for (*W*, *H*) corresponding to converged modes are used as the training dataset for the FCNN model. The framework of the FCNN model includes two subnetworks, namely FCNN-1 and FCNN-2 (as detailed in Section 1 of Supplementary Materials). Once trained, FCNN-1 can determine whether the TE or TM modes converge for a much larger test dataset containing high-resolution (*W'*, *H'*), and FCNN-2 can predict the corresponding polarizer figure of merits *FOM*'s for the converged modes. For clarity in comparison, in our following discussion the parameters of the training and test datasets are labeled in the same manner but with slight difference. For example, *FOM*, *W*, and *H* refer to the parameters for the training dataset, whereas *FOM'*, *W'*, and *H'* correspond to those for the test dataset.

In order to optimize device structural parameters and achieve the maximum polarizer *FOM*, conventional approaches rely on commercial mode simulation software (such as COMSOL Multiphysics and Lumerical FDTD) to carry out exhaustive parameter sweeps over all sets of (*W*, *H*). For instance, scanning $W \in [300, 1000]$ nm and $H \in [100, 300]$ nm at 1-nm resolution would require simulations for over 140,000 sets of (*W*, *H*), with each simulation typically taking 5–7 minutes. As a result, the computing time and cost become extremely high, particularly given the fact that the 2D material films (with thicknesses typically on the order of 1 nm) require ultra-fine mesh resolution to ensure accurate mode simulations.

In contrast, our approach uses machine learning to extract and model the relationships between structural parameters and modal behavior from a small number of (*W*, *H*) sets, allowing for rapid prediction of arbitrary high-resolution (*W'*, *H'*). In addition, performing an exhaustive sweep over all high-resolution (*W'*, *H'*) using our method adds minimal extra time compared to predicting a single set. For example, predicting the *FOM'* for one set with converged mode takes 40 – 80 ms, whereas sweeping over 140,000 sets of (*W'*, *H'*) requires only 25 – 35 seconds, with each additional set contributing less than 1 ms to the total computing time. Except for saving substantial computing time and cost, our method also delivers high prediction accuracy. For example, using 396 sets of (*W*, *H*) as the training dataset, our method can achieve a high accuracy of ~99.0% in predicting mode convergence and a low average deviation (*AD*) of 0.018 when predicting the *FOM'* for over 140,000 sets of (*W'*, *H'*).

In this work, we develop a computer-aided design (CAD) approach for 2D-material-based integrated optical polarizers based on a commonly used FCNN framework in ML. The neural network in our work is implemented based on electronic software and serves as a CAD tool, which is fundamentally different from previous studies that employ photonic hardware to implement optical neural networks for accelerated AI computing [7, 57, 58]. In recent years, ML has been increasingly used in the design of optical devices, such as metasurfaces [13, 59, 60], modulators [61-63], and photodetectors [20, 64, 65], where it is typically employed to map design parameters to device performance for fast performance prediction, parameter optimization, or inverse design – consistent with the strategy adopted in this study. In our work, the FCNN framework consisting of two subnetworks performs more complex ML functions tailored to the specific requirements of designing 2D-material-based optical polarizers, including both performance prediction and mode convergence identification. Specifically, FCNN-1 first identifies physically meaningful regions of the global parameter space, enabling subsequent performance prediction based on FCNN-2 to be performed within this constrained space.

Based on the FCNN framework detailed in Section 1 of Supplementary Materials, we trained models for optimizing the *FOM* of GO-coated Si waveguide polarizers across varying waveguide width *W* and height *H*. For comparison, we selected training datasets containing different numbers of (*W*, *H*) sets, which correspond to different step sizes (*i.e.*, $\Delta$ = 80 nm, 40 nm, and 20 nm) between adjacent parameters within the ranges of *W* $\in$ [300, 1000] nm and *H* $\in$ [100, 300] nm. A smaller step size $\Delta$ results in a larger number of (*W*, *H*) sets in the training dataset. For example, for mode-convergence identification, the dataset with $\Delta$ = 20 nm contains 396 sets, whereas the dataset with $\Delta$ = 80 nm only has 27 sets. The ranges of *W* and *H* were chosen to roughly span the convergence boundaries of the fundamental TE and TM modes for single-mode Si waveguides operating near 1550 nm. In addition, a high-resolution test dataset, generated with $\Delta$' = 1 nm within the same ranges of *W*' $\in$ [300, 1000] nm and *H*' $\in$ [100, 300] nm, was used to evaluate the trained models for mode-convergence identification and polarizer *FOM*' prediction.

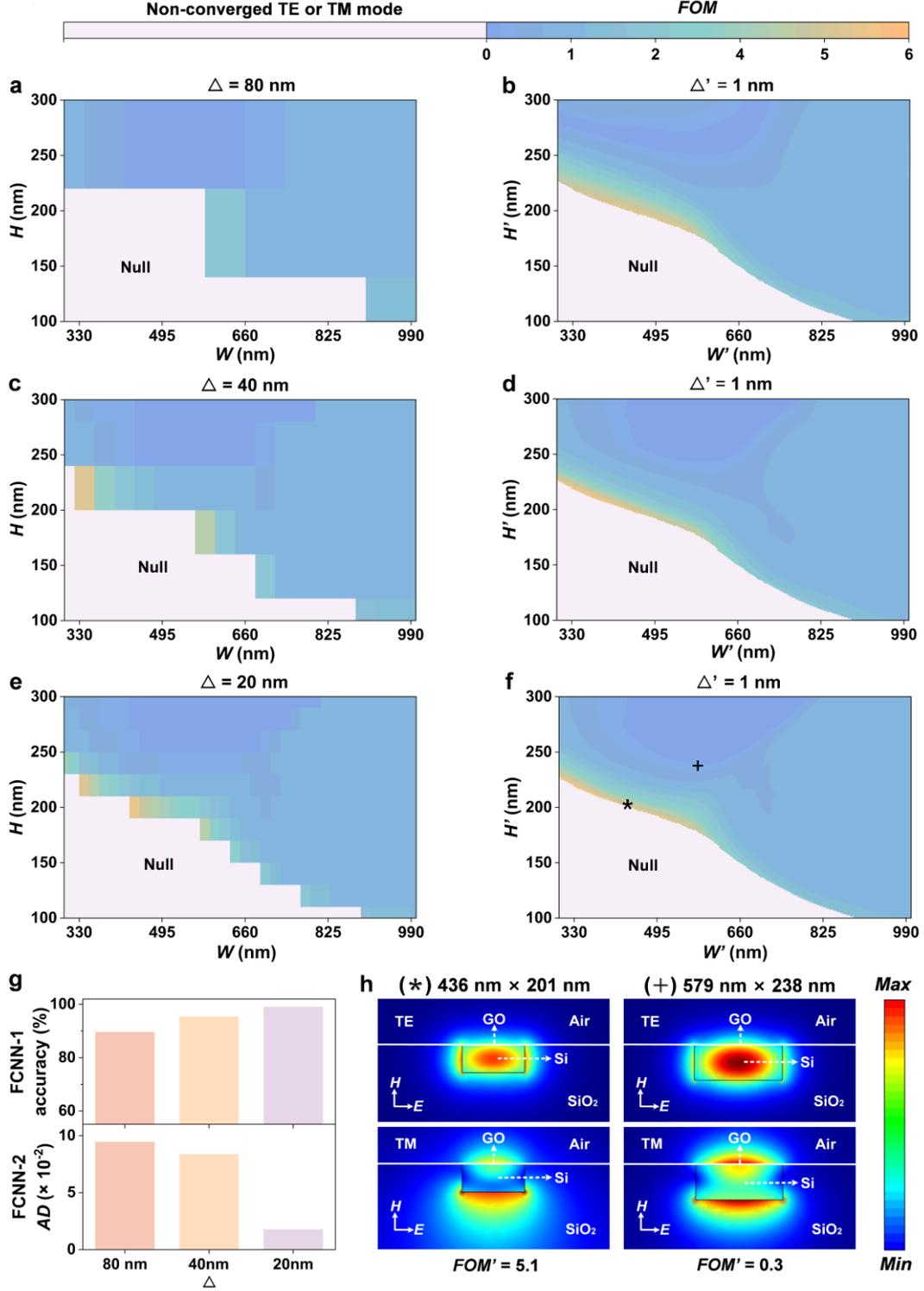

**Fig. 2** | **a, c, e,** *FOM* of GO-Si waveguide polarizer versus low-resolution ($W$, $H$) with $\Delta$ = 80 nm, 40 nm, and 20 nm, respectively, where $\Delta$ is the step size between adjacent waveguide parameters within ranges of $W \in$ [300, 1000] nm and $H \in$ [100, 300] nm. The *FOM* values were calculated based on mode simulations and the 'Null' regions denote the cases of non-converged TE or TM modes in the simulations. **b, d, f,** *FOM'* versus high-resolution ($W'$, $H'$) with $\Delta'$ = 1 nm. The *FOM'* values were predicted using FCNN-2, which was trained by using the data in **a**, **c, e,** respectively. The 'Null' regions denote the cases of non-converged TE or TM modes, which were predicted using FCNN-1. **g,** Accuracy of FCNN-1 and average deviation (*AD*) of FCNN-2 versus $\Delta$. **h,** TE and TM mode profiles corresponding to a high- and a low-*FOM'* point in **f**, marked by '∗' and '+', respectively.

Fig. 2a shows the *FOM* of GO-Si waveguide polarizer versus low-resolution (*W*, *H*) with a step size of Δ = 80 nm. The *FOM* values were calculated based on mode simulations using commercial software (COMSOL Multiphysics), and the 'Null' region corresponds to the (*W*, *H*) with non-converged TE or TM modes in the simulations. In our mode simulations, the GO film parameters (*n*, *k*, *d*) were obtained from our experimental measurements reported in Refs. [30, 32]. For Δ = 80 nm, the (*W*, *H*) parameter space contains only 27 sets. As a result, Fig. 2a appears as large, discretized patches, reflecting the limited resolution of the training dataset.

Fig. 2b shows *FOM*' versus high-resolution (*W*', *H*') with Δ' = 1 nm, which was generated by our FCNN model, trained using the low-resolution dataset in Fig. 2a. The 'Null' region indicates cases of non-converged TE or TM modes identified by FCNN-1, whereas the *FOM*' values in the convergent region were predicted by FCNN-2. Compared with Fig. 2a, the results in Fig. 2b exhibit much higher resolution, where the convergence boundaries and the trend of *FOM*' variation remain consistent with the low-resolution results obtained from mode simulations. This demonstrates that the trained FCNN model can effectively predict high-resolution *FOM*' by capturing how the variations in waveguide structural parameters influence mode convergence and *FOM*. We also note that the change of *FOM*' with either *W*' or *H*' is non-monotonic, mainly resulting from non-monotonic changes in the mode overlap with the GO film. This complex dependence cannot be accurately captured by simple interpolation or polynomial fitting, as these methods inherently assume smooth and globally varying relationships and therefore fail to represent the strongly nonlinear and locally varying behavior of the *FOM* with respect to the waveguide geometry. In contrast, our ML approach does not rely on predefined functional forms, but learns the nonlinear mapping directly from data, enabling it to efficiently capture complex dependencies from low-resolution datasets and thereby facilitate effective optimization of device structural parameters to achieve optimized *FOM'*.

Figs. 2c and 2e show the *FOM* of GO-Si waveguide polarizer versus (*W*, *H*) with step sizes of Δ = 40 nm and 20 nm, respectively. Figs. 2d and 2f show the corresponding *FOM'* versus high-resolution (*W'*, *H'*) with Δ' = 1 nm. For Δ = 40 nm and 20 nm, the training datasets contain 108 and 396 sets of (*W*, *H*), respectively. Compared with Fig. 2a, these larger datasets provide more detailed mode simulation results for training our FCNN model. Similar to Fig. 2b, the higher-resolution results in Figs. 2d and 2f also capture the variation trends of the convergence boundaries and *FOM* obtained from mode simulations, further demonstrating that the FCNN model can effectively perform predictions based on low-resolution datasets.

Although the results in Figs. 2b, 2d, and 2f display the same resolution for (*W'*, *H'*), they actually have different prediction accuracies resulting from different sizes of their training datasets. To quantitatively analyze the prediction accuracy, we plot the accuracy of FCNN-1 and the average deviation (*AD*) of FCNN-2 for various Δ in Fig. 2g. The accuracy of FCNN-1 is defined as the proportion of correctly classified cases, whereas the *AD* of FCNN-2 represents the mean absolute deviation between the values of predicted *FOM'* and simulated *FOM*. As Δ decreases, the training dataset becomes larger, leading to increased accuracy of FCNN-1 and decreased *AD* of FCNN-2. This indicates that a larger training dataset allows the FCNN model to better capture the underlying relationships and thus improve prediction accuracy. At Δ = 20 nm, a maximum FCNN-1 accuracy of ~99.0% and the minimum FCNN-2 *AD* of ~0.018 are achieved. When Δ decreases from 40 nm to 20 nm, the variations in these two parameters become more significant than those observed when Δ decreases from 80 nm to 40 nm. This reflects the fact that the variation in prediction accuracy with the size of training dataset does not follow a linear trend. This behavior originates from the interplay of two competing effects. On one hand, reducing Δ increases the number of training (*W*, *H*) samples, which provides more information for model training to improve prediction accuracy. On the other hand, a smaller Δ also reveals finer variations in the (*W*, *H*) – *FOM* mapping and introduces a larger fraction of samples located in rapid-variation regions, where small geometric changes lead to pronounced *FOM* fluctuations. These fine-scale nonlinear features increase the intrinsic complexity of the learning task, which can counteract the benefit of increased data volume and result in nonlinear changes in the prediction accuracy.

Under a fixed training dataset condition with Δ = 20 nm, we investigated how the configuration of the FCNN framework influences the prediction performance. Taking FCNN-1 as an example, we first tested a two-hidden-layer neural network configuration (32, 64), achieving a classification accuracy of 78.3%. The accuracy increased to ~99.0% for a three-hidden-layer configuration (32, 64, 32), and further to ~99.3% for a four-hidden-layer configuration (32, 64, 64, 32). When the number of the hidden layers was fixed at three, the configuration (16, 32, 16) had a classification accuracy of ~97.1%, whereas a larger-scale configuration (64, 128, 64) with more neurons in each layer yielded a slightly increased classification accuracy of ~99.1%. Considering the above testing results, a three-hidden-layer configuration (32, 64, 32) is chosen for our subsequent analysis to balance the trade-off between improving accuracy and reducing training cost.

Fig. 2h shows the TE and TM mode profiles corresponding to a high- and a low-*FOM'* point in Fig. 2f marked by '✶' and '+', respectively. The *FOM'* values of these two points are ~5.16 and ~0.25, and the corresponding waveguide structural parameters (*W'*, *H'*) are (436 nm, 201 nm) and (579 nm, 238 nm), respectively. The huge difference between the high and low

*FOM'* values highlights the significance of optimizing waveguide structural parameters in the design of 2D-material-based optical polarizers. In addition, *FOM* values of ~5.13 and ~0.25 were obtained from mode simulations using the same waveguide structural parameters, closely matching the *FOM'* values predicted by our FCNN model and reflecting the accuracy of our ML approach. We also note that the highest *FOM'* values in Fig. 2f are achieved near the mode convergence boundary. Although selecting (*W*, *H*) exactly at the convergence boundary can maximize the polarizer *FOM*, this comes at the expense of a limited operation bandwidth. For instance, at (439 nm, 200 nm) near the mode convergence boundary, the *FOM'* value reaches ~5.28, but the associated upper limit of mode convergence bandwidth is only ~1560 nm. For practical devices, the trade-off between an increased polarizer *FOM* and a decreased operation bandwidth should be balanced. A practical solution is to choose (*W*, *H*) slightly offset from the convergence boundary, which can provide a relatively high *FOM* together with a minor decrease in the operation bandwidth. For instance, for geometries slightly offset from the mode convergence boundary, as indicated by the high-*FOM'* point ('✶') in Fig. 2h, the *FOM'* at 1550 nm decreases by only ~0.12, but the corresponding upper limit of mode convergence bandwidth increases to ~1600 nm. These results further reflect the complexity in optimizing the performance of 2D-material-based optical polarizers. Generally, the required operation bandwidths of optical polarizers strongly depend on the intended applications. For example, for optical communications a bandwidth of 100 nm centered at 1550 nm is usually sufficient, whereas optical sensing applications may require a much broader operational bandwidth of up to several hundreds of nanometers. If the target operation wavelength range is specified, our FCNN model can rapidly sweep the entire parameter space to identify the point that satisfies mode convergence in this wavelength range and yields the highest *FOM* value, thereby achieving more accurate design of 2D-material-based optical polarizers.

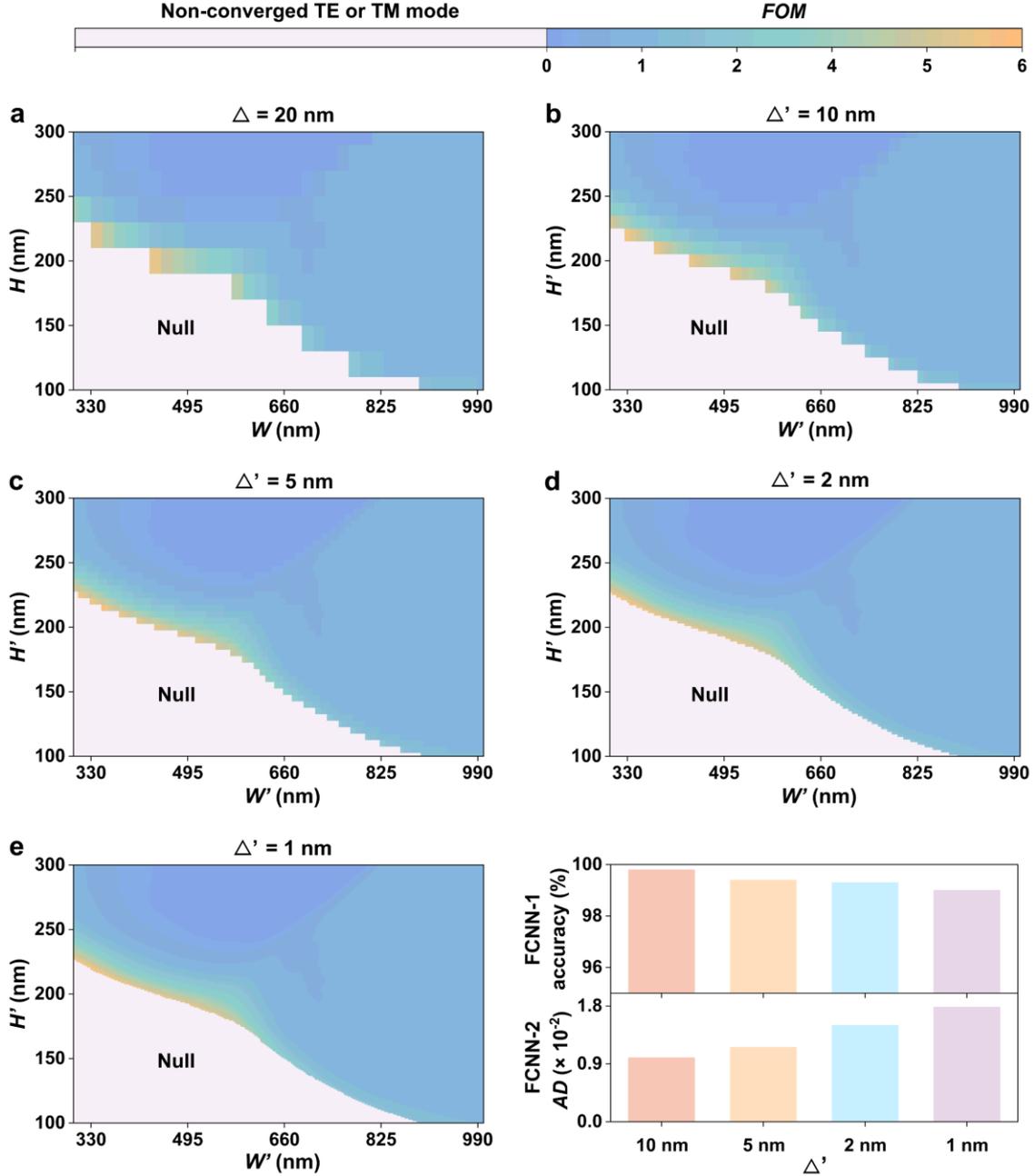

**Fig. 3** | **a,** *FOM* of GO-Si waveguide polarizer versus low-resolution (*W*, *H*) with Δ = 20 nm. The *FOM* values were calculated based on mode simulations and the 'Null' regions denote the cases of non-converged TE or TM modes in the simulations. **b**, **c**, **d**, **e**, *FOM*' versus high-resolution (*W'*, *H'*) with Δ' = 10 nm, 5 nm, 2 nm, 1 nm, respectively. The *FOM*' values were predicted using FCNN-2, which was trained by using the data in **a**. The 'Null' regions represent the cases of non-converged TE or TM modes, which were predicted using FCNN-1. **f,** Accuracy of FCNN-1 and *AD* of FCNN-2 versus Δ'.

In Fig. 2, we compare the performance of our FCNN model using training datasets of different sizes (*i.e.*, Δ = 80, 40, and 20 nm) against a fixed test dataset size of Δ' = 1 nm. In Fig. 3, we further compare the performance of our FCNN model for test datasets of different sizes (*i.e.*, Δ' = 10, 5, 2, and 1 nm) keeping the size of training dataset fixed at Δ = 20 nm. Here, we choose Δ = 20 nm for the training dataset because it provides the best prediction accuracy in Fig. 2.

As shown in Fig. 3a, the training dataset is generated with a step size of Δ = 20 nm and includes the *FOM* values of GO-coated Si waveguide polarizers for 396 sets of (*W*, *H*). Similar to Fig. 2e, the *FOM* values were calculated based on mode simulations, and the 'Null' region corresponds to the (*W*, *H*) with non-converged TE or TM modes. Figs. 3b – 3e show *FOM*' versus higher-resolution (*W*', *H*') with Δ' = 10, 5, 2, and 1 nm, respectively, which were obtained by training the FCNN model using the low-resolution dataset in Fig. 3a. As Δ' decreases from 10 nm to 1 nm, the number of (*W*', *H*') sets in the test dataset increases from 1491 to 140,901, resulting in more refined convergence boundaries and more detailed *FOM*' characteristics.

In Fig. 3f, we compare the accuracy of FCNN-1 and the *AD* of FCNN-2 for various Δ' in Figs. 3b – 3e. A decrease in Δ' (*i.e.*, a larger test dataset) results in decreased accuracy of FCNN-1 and increased *AD* of FCNN-2, showing a trend opposite to that observed in Fig. 2g. This indicates that the FCNN model achieves lower prediction accuracy when applied to finer resolutions of the waveguide structural parameters. When Δ' decreases from 10 nm to 1 nm, the accuracy of FCNN-1 decreases from ~99.8% to ~99.0%, whereas the *AD* of FCNN-2 increases from ~0.01 to ~0.018, showing only minor degradation in the prediction accuracy.

This confirms the accuracy of our FCNN model when applied to test datasets much larger than the training dataset.

To demonstrate the universality of our approach, the FCNN model was also applied to optimizing the *FOM* of $MoS_2$-Si waveguide polarizers, as detailed in Section 2 of Supplementary Materials. Similar to Figs. 2e and 2f, with the training and test step sizes fixed at Δ = 20 nm and Δ' = 1 nm, respectively, the *AD* of FCNN-2 for predicting $MoS_2$-Si polarizers is ~0.033, which is slightly higher than ~0.018 obtained for GO-Si polarizers. Such difference can be attributed to a larger *FOM* value range of $MoS_2$-Si polarizers (*i.e.*, [0, 8]) as compared to that for GO-Si polarizers (*i.e.*, [0, 6]). When the target variable spans a wider range, the FCNN model needs to learn a more extensive input-output mapping. With a fixed training dataset size, certain subranges might receive insufficient coverage, thus leading to increased prediction deviations. Nevertheless, the two types of polarizers exhibit small *AD* values on the order of $10^{-2}$. This confirms the high accuracy of our ML approach when applied to different types of 2D-material-based optical polarizers.

To provide a comprehensive evaluation of our approach, we test our FCNN model under various conditions and compare the performance with respect to computing time and prediction accuracy in Fig. 4. Fig. 4a compares the single-point computing times of the mode simulation and ML methods. Results for ten test points are shown, each corresponding to a specific (*W*, *H*) for mode simulation or a (*W'*, *H'*) serving as input for the FCNN model. The computing times for the FCNN-based ML method are evaluated in two stages, one corresponds to predictions of mode convergence using only FCNN-1, and the other corresponds to predictions of *FOM'* using both FCNN-1 and FCNN-2.

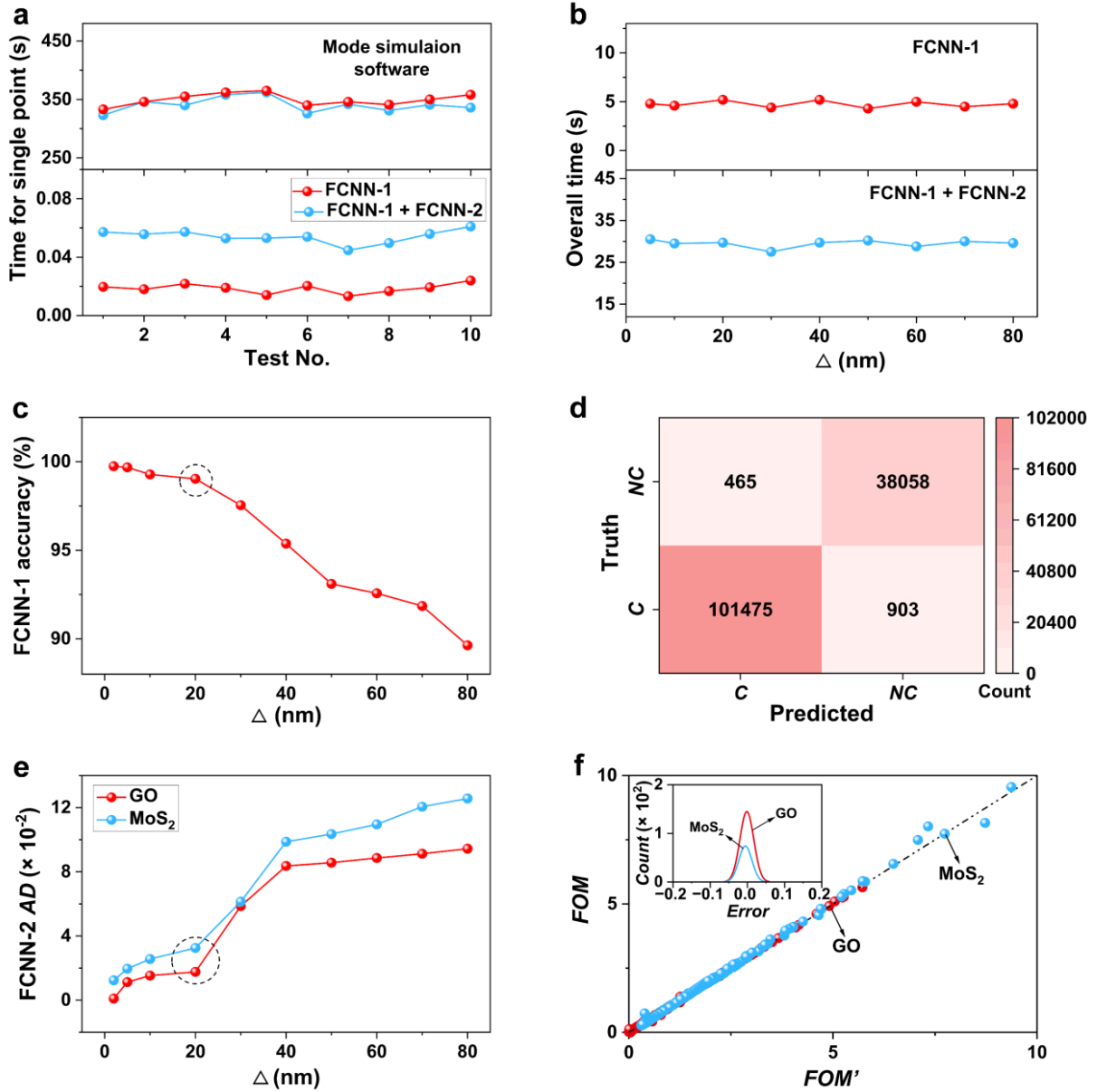

**Fig. 4 | a,** Comparison of single-point computing times (across 10 tests) between the mode simulation and the ML methods. **b**, Overall computing time of the FCNN model for all ($W'$, $H'$) in the test dataset with a step size of $\Delta'$ = 1 nm versus step size of $\Delta$ for ($W$, $H$) in the training dataset. In **a** and **b**, the results for the FCNN model include two curves: one corresponds to predictions of mode convergence using only FCNN-1, the other corresponds to predictions of $FOM'$ using both FCNN-1 and FCNN-2. **c,** Accuracy of FCNN-1 versus $\Delta$. **d**, Confusion matrix of FCNN-1 corresponding to the point in the dashed circle marked in **c**. **e,** $AD$ of FCNN-2 versus $\Delta$ for GO-Si and MoS$_2$-Si polarizers. **f**, Comparison between FCNN-2 predictions and mode simulation results for the points in the dashed circle marked in **e**. Each dot represents one data point, and the black dashed line indicates the diagonal reference.

As can be seen, the mode simulation method requires ~300 – 400 s to finish computing for one point. This process includes three steps: (i) performing mode simulations with software, (ii) manually identifying fundamental TE or TM modes and verifying convergence, and (iii) calculating the *FOM* values based on the mode simulation results. In our case, mode simulations using COMSOL Multiphysics in the first step account for over 80% of the total computing time and were carried out on a standard computer equipped with an Intel(R) Core (TM) i7-7700K CPU running at 4.20 GHz and 32.0 GB of RAM. In contrast, the trained FCNN model completed mode convergence identification through FCNN-1 in only ~0.01 – 0.03 s, and the entire *FOM'* prediction process involving both FCNN-1 and FCNN-2 required only ~0.04 – 0.08 s – about 4 orders of magnitude faster than the mode simulation method. For comparison, all computational stages of our FCNN method were carried out on the same computer as that used for the mode simulation method. The remarkable acceleration of our FCNN method results from replacing repeated numerical solving with a simple forward pass of the trained network, in which the input parameters are processed through a series of matrix multiplications and nonlinear activation functions to produce the output. We also tested the computing time of the mode simulation method and our FCNN method using a low-end computer equipped with 11th Gen Intel(R) Core (TM) i5-1135G7 @ 2.40 GHz and 16.0 GB of RAM. The former needed ~660 – 820 s to finish computing for one point in the training dataset, which is much higher than that on the high-end computer. The latter completed the *FOM'* prediction process involving both FCNN-1 and FCNN-2 in ~0.12 – 0.17 s, with only minimal degradation compared to that on the high-end computer. These results indicate that, unlike the mode simulation method, which is highly dependent on computing resources, our FCNN method exhibits substantially reduced hardware dependence, providing another attractive advantage.

Fig. 4b shows the overall computing time of the FCNN model versus step size of Δ for (*W*, *H*) in the training dataset. The overall computing time refers to the total time required for all (*W'*, *H'*) in the test dataset with a step size of Δ' = 1 nm. Similar to Fig. 2, a larger Δ leads to a smaller size of the training dataset. As the step size Δ increases from 5 nm to 80 nm, the number of (*W*, *H*) sets decreases from 5781 to 27. Since the prediction of FCNN-1 for mode convergence identification mainly depends on the model architecture and the size of test dataset, the overall computing time for FCNN-1 remains largely unaffected by the size of the training dataset, staying nearly constant at ~5 s. On the basis of FCNN-1, FCNN-2 was employed to predict the *FOM'* values for (*W'*, *H'*) corresponding to the converged modes. The overall computing times for both FCNN-1 and FCNN-2 remain in the range of ~25 – 35 s. Compared with FCNN-1, FCNN-2 dominates the overall computing time, accounting for ~80 %. The higher computing time of FCNN-2 arises from its need to perform more complex regression tasks for accurate prediction of *FOM'* values (as compared to FCNN-1 that only carries out binary classification for mode convergence identification), as well as its larger network architecture (*e.g.*, having 128 neurons per hidden layer compared with 64 in FCNN-1). We also note that the overall computing times for both FCNN-1 and FCNN-2 show no significant variation with the size of training dataset, mainly because the forward computing time of the FCNN model primarily depends on the complexity of the model architecture and the size of test dataset.

According to Figs. 4a and 4b, the mode simulation method requires ~300 – 400 s to finish computing for a single point. Exhaustively sweeping all (*W'*, *H'*) in the test dataset with a step size of Δ' = 1 nm results in an overall computing time on the order of ~$10^7$ s (*i.e.*, over 100 days of continuous running), highlighting the extremely demanding computing time and cost. In contrast, our FCNN model offers substantial savings in computing time and cost, requiring less than 40s to exhaustively sweep all (*W'*, *H'*) in a test dataset of the same size. This also highlights the advantage of our FCNN model in handling massive sets of device structural parameters for performance optimization.

In addition to computing time, prediction accuracy serves as another key metric for evaluating the performance of the FCNN model. Fig. 4c illustrates the accuracy of FCNN-1 versus Δ, where the step size of test dataset is fixed at Δ' = 1 nm. Here we present additional results for different Δ values to extend the results in Fig. 2g and provide a more detailed analysis of the influence of training dataset size on the prediction accuracy. The accuracy of FCNN-1 decreases as Δ increases. The decrease is gradual for Δ in the range of [2, 20] nm, and becomes more significant for larger Δ values, indicating that prediction accuracy does not decrease linearly with increasing training dataset size. As a result, important nonlinear features are insufficiently represented in the training data, leading to reduced data diversity and an inadequate representation of critical feature distributions. Subtle variations in the input-output relationship are not sufficiently resolved, which limits the achievable prediction accuracy. This is a combined result caused by multiple factors – most notably reduced data diversity and inadequate representation of feature distributions – which limit FCNN-1 to capture finer variations in the input-output mapping that are essential for achieving higher prediction accuracy.

Although a larger training dataset improves the accuracy of FCNN-1, it also increases the computing time and cost required for constructing the training dataset. This reveals an inherent trade-off between achieving high prediction accuracy and low cost for constructing the training dataset. As the step size Δ decreases, the training dataset grows in size, resulting in increased computing time and cost required for its construction. In our case, the dataset contains 27, 396, and 140,901 samples for Δ = 80 nm, 20 nm, and 1 nm, respectively. The latter two correspond to 14.7 and 5.2 × $10^3$ times increases in the dataset construction time relative to that for Δ = 80 nm. This, together with the fact that further reducing Δ when it is < 20 nm results in only minimal accuracy improvement, leads to our selection of Δ = 20 nm for *FOM* optimization of GO-Si and $MoS_2$-Si polarizers. Generally, the above trade-off can be flexibly managed by tailoring the dataset construction strategy to specific requirements in practical applications. For example, a smaller dataset (*i.e.,* a larger step size) can be used for rapid scanning of the parameter space, improving efficiency and reducing computing time and cost. In contrast, a larger data size in critical performance regions can be employed to enhance prediction accuracy, thereby enabling a more optimized design scheme.

Fig. 4d shows the confusion matrix of FCNN-1 with a step size of Δ = 20 nm for the training dataset, which corresponds to the point in the dashed circle marked in Fig. 4c. The confusion matrix can be used to evaluate the agreement between the truth and predicted labels for mode convergence identification, where '*C*' denotes converged modes and '*NC*' represents non-converged TE or TM modes. The values in the matrix represent the numbers of (*W'*, *H'*) sets for different cases, and their sum equals 140,901, which is the total number of (*W'*, *H'*) sets in the test dataset. In our test, we observed that 101,475 sets for converged modes and 38,058 sets for non-converged modes were correctly identified, and only 465 sets for non-converged modes and 903 sets for converged modes were misclassified as the opposite label. These results in an overall accuracy of FCNN-1 exceeding 99% and an error rate below 1%. This demonstrates that the high accuracy of FCNN-1 for mode convergence identification and its reliability in supporting subsequent prediction of *FOM'* using FCNN-2.

In Figs. 4e and 4f, we further evaluate the prediction accuracy of FCNN-2. Fig. 4e shows the *AD* of FCNN-2 versus step size Δ of the training dataset. Here we show the results for both GO-Si and MoS$_2$-Si polarizers with the same step size of Δ' = 1 nm for the test dataset. Similar to Fig. 4c, the prediction accuracy decreases as Δ increases, and the variation of *AD* with Δ exhibits a nonlinear trend. In addition, FCNN-2 also involves a trade-off between improving prediction accuracy and controlling the cost for constructing the training dataset. As shown in Fig. 4e, the *AD* of FCNN-2 increases noticeably when Δ exceeds 20 nm, whereas further reducing Δ when it is < 20 nm leads to a substantial increase in the cost of training dataset construction, indicating that Δ = 20 nm represents a balanced trade-off.

Fig. 4f shows the comparison between FCNN-2 predictions and mode simulation results for the points in the dashed circle marked in Fig. 4e. The step size of training dataset is Δ = 20 nm, and the step size of test dataset is Δ' = 1 nm. Each dot represents one data point, and the black dashed line indicates the diagonal reference. As can be seen, the points for both GO-Si and MoS$_2$-Si polarizers cluster closely around the diagonal reference, indicating high prediction accuracy of FCNN-2. In addition, the MoS$_2$-Si polarizers exhibit greater errors compared with the GO-Si polarizers. The inset of Fig. 4f shows the corresponding prediction error distribution curves. Compared with the MoS$_2$-Si polarizers, the GO-Si polarizers exhibit a higher and narrower zero-centered error distribution, indicating better prediction accuracy and consistency.

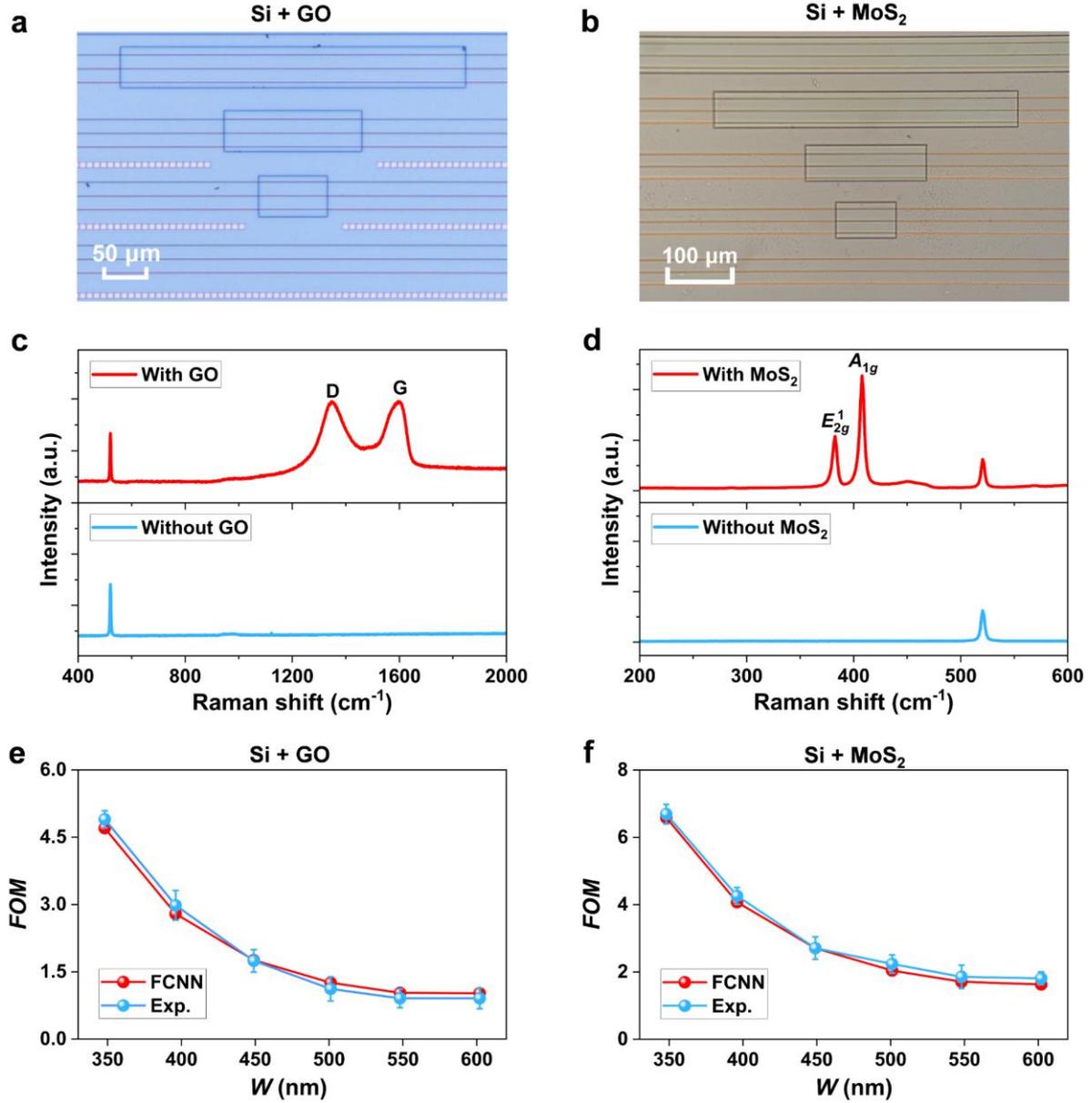

**Fig. 5** | **a, b,** Microscopic images of fabricated GO-Si and MoS$_2$-Si waveguide polarizers on silicon-on-insulator (SOI) chips, respectively. **c, d,** Measured Raman spectra of SOI chips before and after coating 2D GO and MoS$_2$ films, respectively. **e, f,** Experimentally measured *FOM* (Exp.) and *FOM'* predicted by the FCNN model (FCNN) versus Si waveguide width (*W*) for the GO-Si and MoS$_2$-Si polarizers, respectively. In **e** and **f**, the Si waveguide height is *H* = 220 nm.

To further validate the effectiveness of our FCNN model, we practically fabricate GO-Si and MoS$_2$-Si waveguide polarizers. In our fabrication, we first fabricated uncoated Si waveguides on a silicon-on-insulator (SOI) wafer with a 220-nm-thick top Si layer and a 2-μm-thick silica layer. The waveguide patterns were defined using 248-nm deep ultraviolet photolithography, followed by inductively coupled plasma (ICP) etching that enabled waveguide formation. Next, a 1.5-μm-thick silica upper cladding layer was deposited on the SOI chip via plasma enhanced chemical vapor deposition (PECVD). Finally, windows of different lengths were opened on the upper cladding through the processes of photolithography and reactive ion etching (RIE) to enable the coating of 2D material films onto the Si waveguides. All the fabricated Si waveguides had the same length of ~3.0 mm, and the lengths of opened windows (*i.e.*, the 2D film coating lengths) ranged between ~0.1 mm and ~2.2 mm.

Figs. 5a and 5b show microscopic images of the fabricated GO-Si and $MoS_2$-Si waveguide polarizers. The transferred 2D material films exhibit good morphology and uniform coverage across the SOI chip surface, confirming the effectiveness and quality of the film coating. Figs. 5c and 5d show Raman spectra of SOI chips before and after coating 2D GO and $MoS_2$ film, respectively. The Raman spectra were characterized by using a ~514-nm excitation laser. The presence of characteristic peaks at ~1345 $cm^{-1}$ and ~1590 $cm^{-1}$ for the GO-coated chip, and at ~384 $cm^{-1}$ and ~404 $cm^{-1}$ for the $MoS_2$-coated chip, provides evidence for successful on-chip integration of these 2D films.

Figs. 5e and 5f compare the experimentally measured *FOM* values with *FOM'* values predicted by the FCNN model. The data points for the experimental results depict the average of measurements on four devices with different 2D film coating lengths (*i.e.*, 0.2 mm, 0.4 mm, 1.0 mm, 2.2 mm), and the error bars illustrate the variations among these devices. As can be seen, the predicted *FOM'* values agree well with the experimental results, with minor discrepancies mainly arising from fabrication-induced variations in the device structural parameters (*e.g.*, 2D film thicknesses). For both types of polarizers, the predicted and measured values differ by less than 0.2, with the minimum deviation reaching 0.01. These results highlight the high accuracy of our FCNN model in guiding practical device design. Given that the operation principle of our FCNN model is universal, it can also be applied to the design of 2D-material-based optical polarizers on other integrated material platforms such as silicon nitride [66] or lithium niobate [67], which will be the subject of our future work.

As discussed in Fig. 2f, the highest *FOM'* values are usually achieved near the mode convergence boundary, where choosing (*W*, *H*) exactly at the boundary can maximize the polarizer *FOM* but limit the operation bandwidth. Therefore, the variation trend of *FOM'*, which provides valuable guidance for balancing the above trade-off, is more important than simply finding the highest *FOM'* points. As discussed in Fig. 2b, our ML method is particularly powerful in mapping the global variation trend of *FOM'* by rapidly sweeping across all sets of device structural parameters. This highlights the strength of our approach in addressing the complex requirements of practical device design. Finally, it is worth noting that the FCNN method is not limited to the design of polarizers as demonstrated in this work, but also has strong potential for application to the design of more sophisticated optical devices with complex structures, such as waveguide resonators [55, 68], nonlinear optical devices [69, 70], photodetectors [71, 72], and metasurfaces [73, 74], which will be the subject of our future work.

**CONCLUSIONS**

In summary, an FCNN-based ML model is developed for design and optimization of 2D-material-based optical polarizers. Trained on mode simulation results for low-resolution structural parameters, the FCNN model can accurately predict the polarizer FOM values for high-resolution structural parameters and rapidly sweep the entire parameter space to map the global variation trend. The performance of the FCNN model is tested using two types of polarizers with 2D GO and $MoS_2$ films. Results show that our ML method can reduce the overall computing time from several months as required for mode simulations to ~30 s, and achieve accurate FOM predictions with an average deviation of less than 0.04 relative to mode simulation results. In addition, the measured FOM values of fabricated polarizers show excellent agreement with those predicted by the FCNN model, with the discrepancies remaining below 0.2. Our work opens new avenues for leveraging AI to facilitate the design and optimization of 2D-material-based optical polarizers.

## MISCELLANEA

### Funding

This work was supported by the Australian Research Council Centre of Excellence in Optical Microcombs for Breakthrough Science (Grant No. CE230100006), the Australian Research Council Discovery Projects Programs (Grant Nos. CE170100026, DP190103186, FT210100806, DP220100020, and DP240100145), Linkage Program (Grant Nos. LP210200345 and LP210100467), the Industrial Transformation Training Centers scheme (Grant No. IC180100005), the Beijing Natural Science Foundation (Grant No. Z180007), the National Natural Science Foundation of China (Grant No. 12404375), and the Innovation Program for Quantum Science and Technology (Grant No. 2021ZD0300703).

### Author contributions

**Rong Wang:** Writing – original draft, Investigation, Visualization, Formal analysis, Data curation, Validation, Software. **Di Jin:** Writing – Conceptualization, review & editing, Investigation, Visualization, Formal analysis, Validation. **Junkai Hu:** Investigation, Visualization, Formal analysis. **Wenbo Liu:** Investigation. **Yuning Zhang:** Investigation, Funding acquisition, Resources. **Irfan H. Abidi:** Funding acquisition, Resources. **Sumeet Walia:** Funding acquisition, Resources. **Baohua Jia:** Funding acquisition, Resources. **Duan Huang:** Funding acquisition, Supervision, Resources. **Jiayang Wu:** Conceptualization, Writing – review & editing, Investigation, Methodology, Supervision, Project administration. **David J. Moss:** Supervision, Funding acquisition, Project administration, Resources.

### Declaration of competing interest

The authors declare no competing interests.